\documentstyle[11pt]{article}

\newcommand{\be}{\begin{equation}}
\newcommand{\ee}{\end{equation}}
\newcommand{\ben}{\begin{eqnarray}}
\newcommand{\een}{\end{eqnarray}}

\newcommand{\no}{\noindent}
\newcommand{\n}{\label}

\begin{document}

\title{Quintessence inhomogeneous cosmology}

\author{
Luis P Chimento\dag\,, 
Alejandro S Jakubi\dag\,
and Diego Pav\'on\ddag\
\\
\\
\dag\ {\small Departamento de F\'{\i}sica, Universidad de 
Buenos Aires, 1428~Buenos Aires, Argentina}\\
\ddag\ {\small Departament de F\'{\i}sica, Universidad Aut\'onoma 
de Barcelona, 08193 Bellaterra, Spain}\\
}

\maketitle
\begin{abstract}
The Einstein--Klein--Gordon field equations
are solved in a inhomogeneous shear--free universe containing a 
material fluid, a self--interacting scalar field, a
variable cosmological term, and a heat flux.
A quintessence--dominated scenario arises with a power--law 
accelerated expansion compatible with the currently observed 
homogeneous universe.
\end{abstract}

\section{Introduction}

Recently, there have been claims in the literature that the Universe, besides
its content in normal matter and radiation, must possess a not yet identified
component (usually called {\em quintessence} matter, Q-matter for short)
\cite{turner1}, \cite{caldwell}, \cite{zlatev}, characterized
by a negative pressure, and possibly a cosmological term. These claims were prompted at the
realization that the clustered matter component can account at most for one
third of the critical density. Therefore, an additional ``soft" (i.e.
non-clustered) component is needed if the critical density predicted by many
inflationary models is to be achieved.

Very often the geometry of the proposed models is very simple, just
Friedmann-Lema\^{\i}tre-Robertson-Walker (FLRW). In constrast to FLRW models,
inhomogeneous spaces are in general compatible with heat fluxes, and these
might imply important consequences such as inflation or the
avoidance of the initial singularity \cite{dadhich}. Here we focus on an
isotropic but inhomogeneous spherically symmetric universe which besides a
material fluid contains a self-interacting scalar field (which can be
interpreted as Q-matter), and a cosmological term, $\Lambda$ which, in
general, may vary with time.

Density inhomogenities triggered by gravitational instability must be present
at any stage of evolution. We only mention that the negative pressure
associated to  Q-matter and $\Lambda$ will tend to slow down the growing modes
(see {\em e.g.} \cite{turner1},  \cite{adz}), and shift the epoch
of matter-radiation equality toward more recent times.
 
\section{Einstein-Klein-Gordon field equations}

Let us consider a shear--free spherically--symmetric spacetime with metric
\cite{nariai}

\begin{equation}\label{ds3}
{ \ ds}^{2}=\frac{1}{{ F}(t, \,r)^{2}}\left[
\,
- v(t,r)^{2}\,{ \ d}\,t^{{ 2\ }}+
 \ d\,r^2  +
  r^2\, \mbox{d}\Omega^2
    \right].
\end{equation}

\no where as usual $\mbox{d}\Omega^2 \equiv \mbox{d}\theta^{2} + \sin^{2}\,
\theta \, \mbox{d} \phi^{2}$. Units have been chosen so that $c = G = 1$. As
sources of the gravitational field we take: a fluid of material energy density
$\rho_f = \rho_f (r,t)$, hydrostatic pressure $p_f = p_f(r,t)$, with a radial
heat flow ($q_{r} = q_{r}(r,t) \, $ and $\, q_{t} = q_{\theta} = q_{\phi} =
0$), plus a cosmological term, related to the energy density of vacuum by
$\Lambda = 8\pi \rho_{vac}$, that depends only on time $\Lambda = \Lambda(t)$,
and a self-interacting scalar field $\phi$ driven by the potential $V(\phi)$
whose equation of state is $p_{\phi}=\left(\gamma_{\phi}-1\right)\rho_{\phi}$.
Hence the scalar field can be interpreted as Q-matter -see e.g.
\cite{zlatev}. The stress energy-tensor of the normal matter, with a heat
flow, plus Q-matter (scalar field) and the cosmological term is

\begin{equation}
T^{i}_{k} = (\rho_f+\rho_{\phi} + p_f+p_{\phi}) u^{i} u_{k} +
(\Lambda - p_f-p_{\phi}) \delta^{i}_{k} + q^{i} u_{k}
+q_{k} u^{i} \, ,
\label{2}
\end{equation}

\no As equation of state for the fluid we choose
$p_f=\left(\gamma_f-1\right)\rho_f$ where $\gamma_{f}$ is a function of $t$
and $r$. Taking into account the additivity of the stress-energy tensor it
makes sense to consider an effective perfect fluid description with equation
of state $p=\left(\gamma-1\right)\rho$ where $p=p_f+p_{\phi}$,
$\rho=\rho_f+\rho_{\phi}$ and

\be
\n{gamma}
\gamma=\frac{\gamma_f\rho_f+\gamma_{\phi}\rho_{\phi}}
{\rho_f+\rho_{\phi}} \, ,
\ee
is the overall (i.e. effective) adiabatic index.
The requirement that the cosmological term $\Lambda$ is just a function of $t$
leads to the restriction that $\gamma$ also depends only on $t$ to render the
system of Einstein equations integrable. The nice result we are seeking is a
solution that has an asymptotic FLRW stage, with $\Lambda $ evolving towards a
constant, and the heat flow vanishing in that limit \cite{lchajdp}.

To write the Einstein equations we use the ansatz $F=a(t) + b(t)x$ and $v=c(t)
+d(t)x$ with the constraints $a(t)\, d(t)=b(t)\, c(t)$. This set of metrics
contains those of Modak ($b=0$) \cite{modak}, Bergmann ($c=a, d=b$)
\cite{bergmann} and Maiti ($b=d=k \, a/4 \, ,$ with $k=0, \pm1$) \cite{maiti}.
Another possibility arises when $d=0$, then re-defining the time by
$v dt\to dt$, the Einstein equations are

\begin{equation} \label{00c}
\rho+\Lambda=12 ab+3\dot a^2+6\dot a\dot b x+3\dot b^2 x^2 ,
\end{equation}

\begin{equation} \label{11c}
p-\Lambda=\left(2b\ddot b-3\dot b^2\right)x^2+
2\left(2b^2-3\dot a\dot b+a\ddot b+\ddot a b\right)x-
8ab-3\dot a^2+2a \ddot{a} \, ,
\end{equation}

\begin{equation} \label{03c}
q_{r} =-4\sqrt{x}\dot b\left(a+bx\right)^{2} ,
\end{equation}

\no where $x=r^2$ and the overdot indicates $\partial/\partial t$.
Imposing that $\Lambda=\Lambda(t)$, the general solution
to these equations has the form

\begin{equation} \label{aw}
a=-2\exp{\left(\int dt \,w\right)}\int dt \,w^2
\int \frac{dt}{w^2} \, , \qquad 
b=\exp{\left(\int dt \,w\right)}\, ,
\end{equation}

\no where $w=2/\int dt (2-3\gamma)$, provided $\gamma\neq 2/3$. Inserting
(\ref{aw}) in (\ref{00c}), (\ref{11c}) and (\ref{03c}) we easily compute 
the cosmological constant and the heat flow. 
Finally the FLRW metric is

\begin{equation} \label{ds5}
ds^2=\frac{1}{\left(1+Mr^2\right)^2}
\left[-d\tau^2+ R^2 \,\left(dr^2+r^2\, d\Omega^2\right)\right]\, ,
\end{equation}

\no where we have introduced the time coordinate $d\tau= dt/a$, $M= b/a \, $
and $R=1/|a|$. This metric is conformal to FLRW, and the conformal factor
approaches unity when $M\to 0$.

\section{Constant adiabatic index}

When $\gamma$ is a constant different from $2/3$, the general solution of
(\ref{00c}) and (\ref{11c}) becomes

\begin{equation} \label{at}
a(t)=C_1 \Delta t^{-\frac{2}{3\gamma-2}}+
C_2 \Delta t^{-\frac{3\gamma}{3\gamma-2}}
-\frac{1}{3}K \Delta t^{6\frac{\gamma-1}{3\gamma-2}} \, ,
\end{equation}

\begin{equation} \label{Mt}
M(t)=K\left(C_1+C_2\Delta t^{-1}-\frac{1}{3}K\Delta t^2\right)^{-1}.
\end{equation}

\no Two alternatives of asymptotically expanding universes appear depending on
the map between $t$ and $\tau$.

\bigskip
\underline{Case $\Delta t\to 0$}

\bigskip

In this limit we obtain

\begin{equation} \label{Ra1}
R(\tau)\simeq \frac{1}{\mid C_{2}\mid}\left[
\frac{2 C_{2} \left(1-3\gamma\right)}{2-3\gamma}
\Delta\tau\right]^{\frac{3\gamma}{2\left(3\gamma-1\right)}} ,
\end{equation}

\begin{equation} \label{lambdaa1}
\Lambda(\tau)\simeq \frac{3\left(2-3\gamma\right)^2}
{4\left(1-3\gamma\right)^2\Delta\tau^2}  ,
\end{equation}

\begin{equation} \label{qa1}
q_{r}(r,\tau)\simeq \frac{8KC_2}{3\gamma-2}\,\,r
\left[\frac{2C_2\left(1-3\gamma\right)}{2-3\gamma}\Delta\tau\right]
^{\frac{9\gamma}{2\left(1-3\gamma\right)}},
\end{equation}

\no for $\gamma\ne 1/3$. When $1/3<\gamma<2/3$ we have, for large cosmological
time $\tau$, an accelerating universe that homogenizes with vanishing
cosmological term and heat flow. In this stage we have a final power-law
expansion era. For $\gamma=1/3$ we have asymptotically a de Sitter 
universe with finite limit cosmological term.
For the remaining values of $\gamma$ the universe begins at a homogeneous
singularity with a divergent cosmological term. When $\gamma<1/3$, the heat
flux asymptotically vanishes near the singularity, while for $\gamma>2/3$ it
diverges.

\bigskip

\underline{Case $\Delta t\to\infty$}

\bigskip

In this limit we obtain

\begin{equation} \label{Ra2}
R(\tau)\simeq -\frac{3}{K}\left[\frac{K\left(4-3\gamma\right)}
{3\left(2-3\gamma\right)}\Delta\tau\right]^
{\frac{6\left(1-\gamma\right)}{4-3\gamma}},
\end{equation}

\begin{equation} \label{lambdaa2}
\Lambda(\tau)\simeq -\frac{24\left(2-3\gamma\right)^2}{\left(4-3\gamma\right)^2
\Delta\tau^2},
\end{equation}

\begin{equation} \label{qa2}
q_{r}(r,\tau)\simeq -24\frac{\left(2-3\gamma\right)^2}
{\left(4-3\gamma\right)^3}\frac{r}{\Delta\tau^3},
\end{equation}

\no for $\gamma\ne 2/3$. When $2/3\le\gamma\le 1$ the universe  homogenizes
for large cosmological time with vanishing cosmological term and heat flow.
When $\gamma=1$, the late time evolution changes to an
asymptotically Minkowski stage. For $1<\gamma<4/3$ the universe starts
homogeneously in the remote past with a vanishing scale factor, cosmological
term and heat flow. For the remaining values of $\gamma$ the universe begins
at a homogeneous singularity with a divergent cosmological term.

An exact solution with explicit dependence on the asymptotic cosmological time
$\tau$ can be found when the integration constants $C_{1}$ and $C_{2}$ vanish.
In such a case the metric is

\begin{equation}
\label{metric}
ds^2=\frac{1}{\left(1+m\Delta\tau^{2\frac{2-3\gamma}{4-3\gamma}}\,r^2\right)^
2}\left[-d\tau^2+\Delta\tau^{12\frac{1-\gamma}{4-3\gamma}}\left(\ d\,r^2  +
  r^2\,d\Omega^2
\right)\right],
\end{equation}

\noindent where $m$ is a redefinition of the old integration constant $K$, the
adiabatic index $\gamma$ and $r_0$. The last constant was introduced by
scaling the radial coordinate $r\to r_{0}\, r$.

\section{Asymptotic evolution to a quintessence--dominated era}

As a first stage of towards more general scenarios with a slowly time--varying
$\gamma$, we will explore a model that evolves towards an asymptotic FLRW
regime dominated by Q--matter (i.e. the scalar field). We will show that this
system approaches to the constant $\gamma$ solutions for large times found
above. In this regime the equations of Einstein-Klein-Gordon become

\begin{equation} \label{00RW}
3H^2\simeq\rho_f+\frac{1}{2}\dot\phi^2+V\left(\phi\right)+\Lambda \, ,
\end{equation}

\begin{equation} \label{KGRW}
\ddot\phi+3 H\dot\phi+ \frac{dV\left(\phi\right)}{d\phi}\simeq 0 ,
\end{equation}

\no where $H=\dot R/R$ and a dot means $d/d\tau$ in this section. In last
section we found that the general asymptotic  solution for the scale factor
$R(\tau)\propto\Delta\tau^\alpha$ has the power--law behaviors (\ref{Ra1}),
(\ref{Ra2}) for any value of the effective adiabatic index $\gamma$. Then,
using these expressions and (\ref{lambdaa1}) and (\ref{lambdaa2}) together
with (\ref{00RW}) and (\ref{KGRW}), we can investigate the asymptotic limit in
which the energy of the scalar field dominates over the contribution of the
perfect fluid. In the regime that $3\alpha\gamma_f>2$ the adiabatic scalar
field index can be approximated by

\be
\n{gammaphi3}
\gamma_ {\phi}\simeq\frac{2}{3\alpha}
\left[1+\left(1-\frac{3\gamma_f}{2}\sigma
\right)\right],
\ee

\no where $\sigma=\rho_f/\rho_{\phi}\ll 1$. Inserting these equations in
(\ref{gamma}) we obtain the first correction to the effective adiabatic index

\be
\n{gamma1}
\gamma\simeq\frac{2}{3}\left[1\pm\sqrt{\frac{11}{12}\left(3\gamma_f-2\right)\sigma}\right]
\ee

\no
The negative branch of (\ref{gamma1}) yields a consistent asymptotic solution
for the range ${\textstyle{1\over 3}}<\gamma<{\textstyle{2\over 3}}$. We note
that this solution describes a deflationary stage with a limiting exponent
$\alpha=1$.

Oftenly power-law evolution of the scale factor is associated with logarithmic
dependence of the scalar field on proper time \cite{Chi98}. Thus, assuming
that $\phi(\tau)\simeq C\ln \tau$ with the constant $C$ to be determined by
the system of equations (\ref{00RW}) and (\ref{KGRW}), and  using these
expressions together with (\ref{lambdaa1}) and (\ref{lambdaa2}) in
(\ref{00RW}) and (\ref{KGRW}) it follows that the leading term of $V(\phi)$
for large $\phi$ is

\begin{equation} \label{Va}
V\left(\phi\right)\simeq V_0 e^{-A\phi}
\end{equation}

\no Using the dominant value of the effective adiabatic index we find $A^2=2$,
$V_0=2$ and $C=1/\sqrt{2}$.

The models considered in this section are based on the notion of ``late time
dominating field" (LTDF), a form of quintessence in which the field $\phi$
rolls down a potential $V(\phi)$ according to an attractor solution to the
equations of motion. The ratio $\sigma$ of the background fluid to the field
energy changes steadily as $\phi$ proceeds down its path. This is desirable
because in that way the Q-matter ultimately dominates the energy density and
drives the universe toward an accelerated expansion \cite{perlmutter1},
\cite{riess}.

\section{Concluding remarks}

We have investigated a class of solutions of the Einstein field equations with
a variable cosmological term, heat flow and a fluid with variable adiabatic
index that includes those of Modak, Bergmann and Maiti and contains a new
exact conformally flat solution. We have found that asymptotically expanding
universes occur when $1/3<\gamma<1$ that homogenizes for large cosmological
time  with vanishing cosmological term and heat flow. For $1/3<\gamma<2/3$ the
evolution is given by (\ref{Ra1}) and corresponds to a power-law accelerated
expansion for large cosmological time $\tau$.
On the other hand, when $2/3\le\gamma<1$ even though an asymptotic negative
cosmological term occurs, the universe evolves toward a decelerated expansion.
The particular case $\gamma=1/3$ leads asymptotically to a de Sitter universe
with a finite limit for $\Lambda$. We have shown that homogeneization occurs
also for a time dependent adiabatic index provided it has a constant limit,
for $t\to 0$ and  $t\to\infty$, repectively, and is analytic about these
points.

We have carried out a detailed analysis of a model in which Q--matter dominates
over cold dark matter. This LTDF solution is an attractor because, even for
large initial inhomogeneities and a wide range of initial values for $\phi$
and $\dot{\phi}$, the evolution approaches a common path. It was shown that
this model can be realized for a wide range of potentials provided they have
an exponential tail.
Our LTDF model only requires that the potential has an asymptotic exponential
shape for large $\phi$. So, the interesting and significant features of our
model are: (a) a wide range of initial conditions are drawn towards a common
evolution; (b) the LTDF maintain some finite difference in the equation of
state such that the field energy  eventually dominates and the universe enters
a period of acceleration.

\section*{Acknowledgments}

This work was partially supported by the Spanish Ministry 
of Education under Grant PB94-0718, and the University of 
Buenos Aires under Grant TX-93.

\end{document}